\documentclass[journal,final,letterpaper]{IEEEtran}
\usepackage{graphicx}
\usepackage{setspace}
\usepackage{url}

\IEEEoverridecommandlockouts

\begin{document}

\newcommand{\longtitle}{A Research Framework for the Clean-Slate
  Design of Next-Generation Optical Access}
\newcommand{\shorttitle}{\longtitle}
\title{\longtitle}

\author{Kyeong Soo Kim, \IEEEmembership{Member, IEEE}
  \thanks{This work was supported in part by Amazon Web Services (AWS) in
    Education Research Grant.}
  \thanks{K. S. Kim is with the College of Engineering, Swansea University,
    Swansea, SA2 8PP, Wales United Kingdom. (e-mail: k.s.kim@swansea.ac.uk).}
}

\markboth{K. S. Kim: \shorttitle}{K. S. Kim: \shorttitle}

\maketitle

\begin{abstract}
  A comprehensive research framework for a comparative analysis of candidate
  network architectures and protocols in the clean-slate design of
  next-generation optical access is proposed. The proposed research framework
  consists of a comparative analysis framework based on multivariate
  non-inferiority testing and a notion of equivalent circuit rate taking into
  account user-perceived performances, and a virtual test bed providing a
  complete experimental platform for the comparative analysis. The capability of
  the research framework is demonstrated through numerical results from the
  study of the elasticity of hybrid TDM/WDM-PON based on tunable transceivers.
\end{abstract}


\section{Introduction}
While investigating the issues of \textit{quality of experience
  (QoE)}, \textit{elasticity}\footnote{It means the ability to manage
  overall performances to a certain level by fast provisioning of
  network resources based on user demands.}, and \textit{energy
  efficiency} in next-generation optical access (NGOA) as part of
research programs since 2009 \cite{Kim:10-1,Kim:10-3,Kim:11-2} with a
major focus on the solutions beyond 10-Gbit/s Ethernet passive optical
network (10G-EPON) and 10-Gigabit-capable PON (XG-PON) (e.g., NG-PON2
\cite{Kani:09} by ITU-T), we have noted that the progress in the
clean-slate design of NGOA is impeded by the absence of a
comprehensive research framework for a comparative analysis of
candidate network architectures and protocols. In fact, many NGOA
network architectures have been proposed by both academia and industry
and are now under extensive study (e.g.,
\cite{Aurzada:11,Grobe:11,Song:10,Lovric:10}). Unfortunately, most of
the existing works lack a systematic comparison of these candidate
architectures under realistic operating environments; they are based
on the comparison of network-level performances (e.g., packet delay
and throughput), reaches, splitting ratios, and energy consumptions
under static or limited statistical traffic configurations without
taking into account the actual performances perceived by end-users
that reflect the impact of higher-layer protocols including
transmission control protocol (TCP) flow and congestion control.

Because of the complexity of protocols and the interactive nature of
traffic involved in the study of network architectures, researchers
now heavily depend on experiments with simulation models or test beds
implementing proposed architectures and relevant protocols, rather
than traditional mathematical analyses under simplifying
assumptions. In this regard, a research framework for the comparative
analysis of NGOA architectures and protocols should specify how to
generate traffic and measure performances during the experiment,
together with a systematic comparison of the measured performances
from the experiments. Note that, due to the shift toward experiments,
comparison procedures should be able to take into account the
statistical variability in measured data from the experiments.

In this paper a new research framework for the clean-slate design of
NGOA architectures and protocols is proposed. The proposed research
framework consists of two major components, i.e., a comparative
analysis framework and a virtual test bed for experiments.

The comparative analysis framework is based on a multivariate non-inferiority
testing procedure \cite{laster03:_non,PST:PST384} and a notion of equivalent
circuit rate (ECR) \cite{shankaranarayanan:01}. In this framework user-perceived
performances of representative services --- including web browsing (i.e.,
hypertext transfer protocol (HTTP)), file downloading (i.e., file transfer
protocol (FTP)), and streaming video (i.e., H.264/advanced video coding (AVC)
with user datagram protocol (UDP)) --- are compared in an integrated way using
statistical hypothesis testing procedures.

The virtual test bed is basically simulation models of the proposed
architectures and protocols. Unlike many existing works in the area of optical
access that mainly focus on the issues up to the data link layer (e.g.,
\cite{Nakagawa:07,Noda:10,Qian:10,Kim:05-1}), the virtual test bed provides a
complete experimental environment with session-level traffic generation (based
on user behaviors) and performance gathering (as measures for user-perceived
performances) as well as models for the whole network protocol stack (including
TCP/Internet protocol (IP)).

The rest of the paper is organized as follows: Section
\ref{sec:comparison_framework} describes the comparative analysis
framework based on the multivariate non-inferiority testing and the
ECR. Section \ref{sec:virtual_test_bed} provides an overview of the
current implementation of the virtual test bed and discusses plans and
strategies for its improvement in the next version. Section
\ref{sec:preliminary_results} presents results from the study of the
elasticity of hybrid time division multiplexing (TDM)/wavelength
division multiplexing (WDM)-PON to demonstrate the capability of the
proposed research framework. Section \ref{sec:conclusion} concludes
the discussions in this paper.

\section{A Comparative Analysis Framework}
\label{sec:comparison_framework}
\begin{figure}[!hbt]
  \centering
  \includegraphics*[angle=-90,width=.9\linewidth]{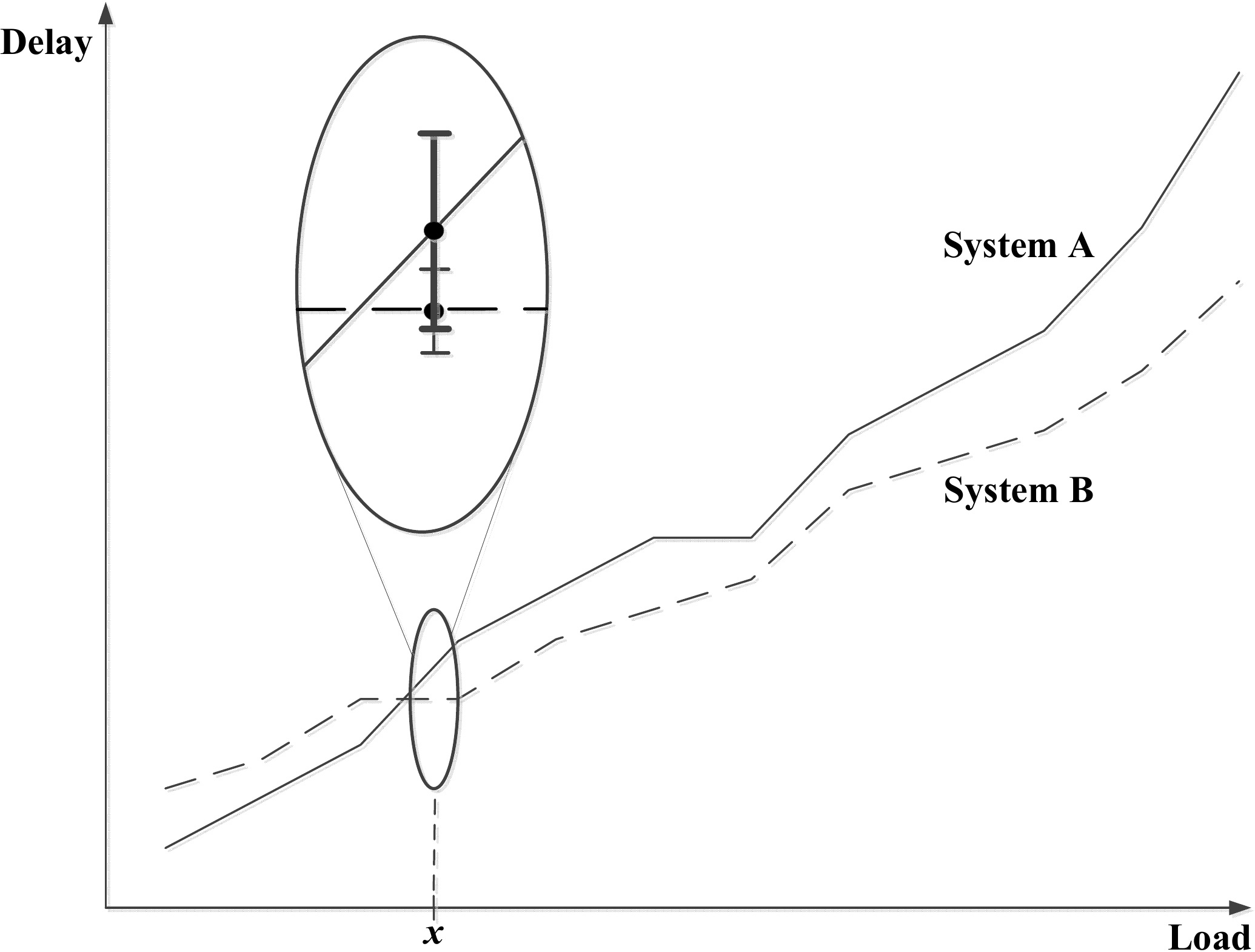}
  \caption{An example of comparison between two delay curves.}
  \label{fig:comparative_example}
\end{figure}

Fig. \ref{fig:comparative_example} illustrates an example of
comparison between two delay curves at a certain value of load (i.e.,
$x$), which is typical in the performance analysis of a new proposed
system with respect to an existing one.
At first glance, the comparison seems straightforward; one can say
from the curves that the system B provides a lower delay than the
system A at load $x$. Once taking into account the statistical
variability in measured data (i.e., the overlapped confidence
intervals in the enlarged part), however, we can find that the
comparison is no longer straightforward and that a statistical
approach is needed in this regard.

Note that most of the works in networking area lack a statistical
approach in the performance comparison and just provide observations
on certain trends. For example, ``the system B provides a better delay
performance than the system A when the load is greater than $x$'' is a
typical observation for Fig. \ref{fig:comparative_example}, and the
exact value of $x$ is not that critical. As we will discuss shortly in
this section, however, the strict comparison of performances at a
certain input value is quite critical to comparison frameworks like
the ECR. Also, considering other performance measures (e.g.,
throughput) as well as delay curves for multiple services (e.g., HTTP
and FTP) altogether, the performance comparison between the systems
becomes even more complicated.

As we are at an early stage of the clean-slate design of NGOA, we
therefore need to establish a new framework for comparing candidate
network architectures and protocols which meets the following
requirements:
\begin{itemize}
\item A comparison procedure should \textit{take into account the
    statistical variability} in measured data resulting from the
  experiments.
\item Multiple performance measures should be compared together
  \textit{in an integrated way}.
\item Measures for the comparison should be \textit{user-oriented}. In
  other words, they should reflect the quality of experience (QoE),
  rather than the quality of service (QoS).
\end{itemize}

To meet these requirements, we have been working on a new comparative
analysis framework based on the non-inferiority testing for the
comparison procedure and the ECR for the quantification of the
resulting performance. The non-inferiority testing is a one-sided
variant of the equivalence testing that is frequently used in Medicine
and Biology for the establishment of the equivalence (often called
\textit{bioequivalence}) between two different clinical trials or
drugs \cite{Berger:96,SIM:SIM985,PST:PST384}. The ECR, on the other
hand, was originally proposed for the quantification of the bandwidth
of hybrid fiber coaxial (HFC) cable-based shared access network with
respect to that of the digital subscriber line (DSL)-based dedicated
access network in terms of web page delay as a measure for
user-perceived performances \cite{shankaranarayanan:01}. The ECR
framework has been extended for a quantitative comparison of optical
access architectures in \cite{Kim:10-1,Kim:10-3}.

Combining these two frameworks, we can meet the aforementioned
requirements: The non-inferiority testing procedure is based on
statistical hypothesis testing and as such takes into account the
statistical variability inherent in measurements as well as
experiments. To compare multiple performance measures in an integrated
way, we extend the non-inferiority testing with the intersection-union
testing (IUT) as described in \cite{Berger:96}. The third requirement
is met by the ECR which enables us to quantify the relative capacity
of a candidate system with respect to a reference one based on
user-perceived performances at the application level.

\begin{figure}[!hbt]
  \centering
  \includegraphics*[angle=-90,width=\linewidth]{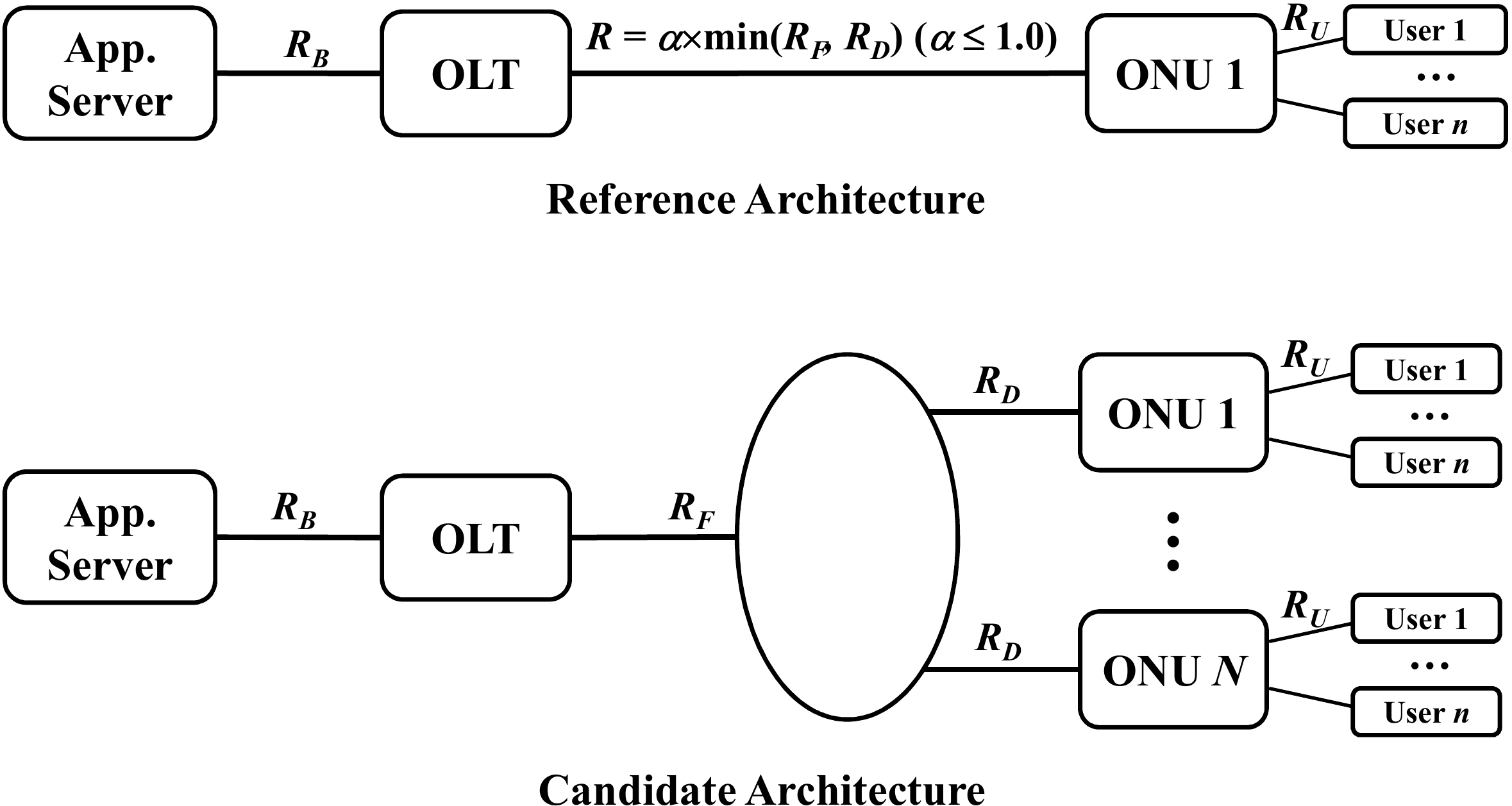}
  \caption{A comparative analysis framework based on the ECR.}
  \label{fig:comparative_analysis_framework}
\end{figure}

Fig. \ref{fig:comparative_analysis_framework} shows the comparative
analysis framework based on the ECR for the investigation of NGOA
architectures and protocols, where $R_{B}$, $R_{F}$, $R_{D}$, and
$R_{U}$ denote backbone, feeder, distribution, and user network
interface (UNI) rates and $R$ the line rate connecting the optical
line terminal (OLT) and the optical network unit (ONU) of a reference
architecture.
In this framework we compare the user-perceived performances of both
the candidate and the reference architectures and find the value of
$R$ (i.e., ECR) for the reference architecture which gives
user-perceived performances statistically equivalent to those of the
candidate architecture. In case of a shared architecture, for
instance, because of contention for a feeder capacity among multiple
ONUs and for a distribution capacity among multiple users connected to
the same ONU, each user's share of capacity cannot be greater than the
minimum of the feeder and the distribution capacities. Therefore the
user-perceived performance would be similar to that of the reference
architecture with the line rate equal to or less than the minimum of
the feeder and the distribution rates of the shared architecture.

\begin{figure*}[!hbt]
  \begin{minipage}{.48\linewidth}
    \begin{center}
      \includegraphics*[angle=-90,width=.8\linewidth]{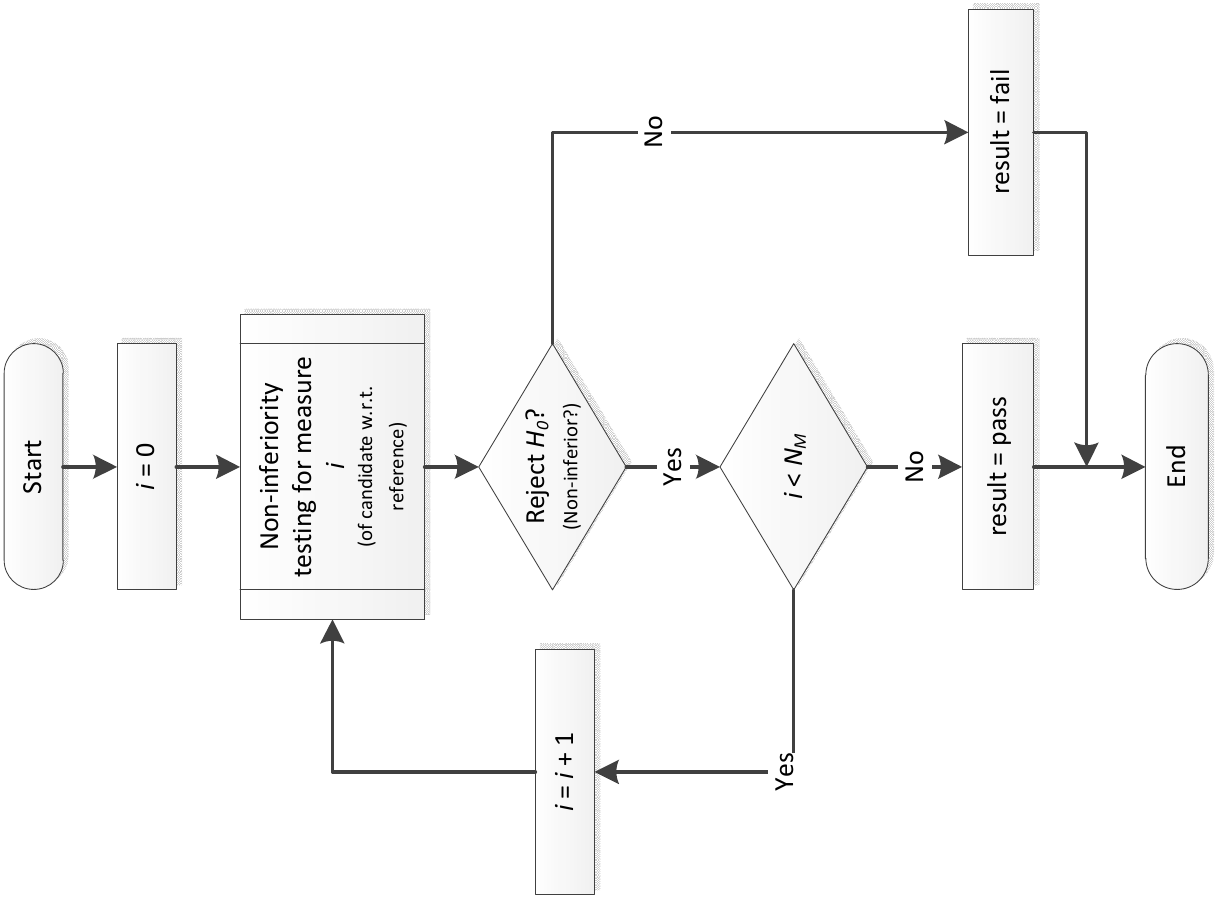}\\
      {\scriptsize (a)}
    \end{center}
  \end{minipage}
  \hfill
  \begin{minipage}{.48\linewidth}
    \begin{center}
      \includegraphics*[angle=-90,width=.8\linewidth]{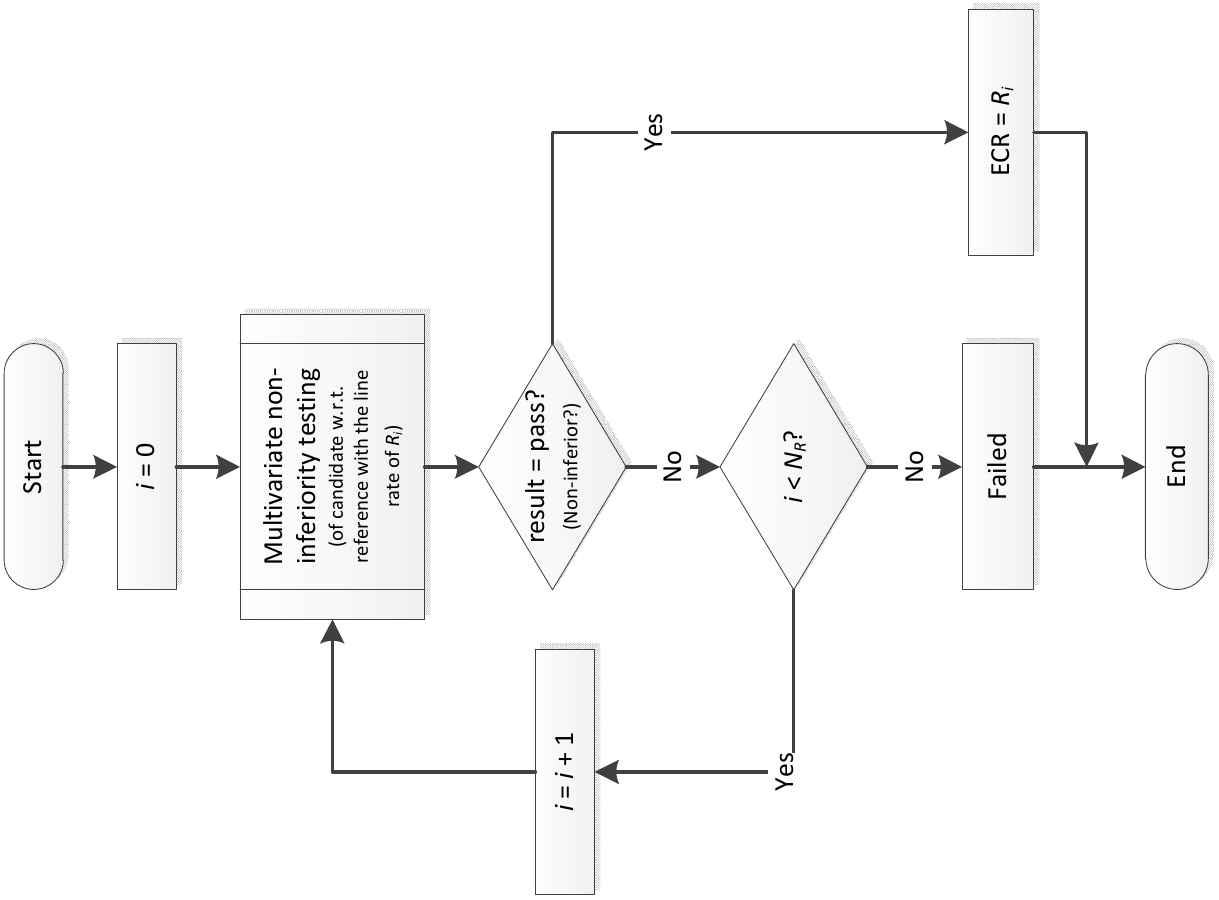}\\
      {\scriptsize (b)}
    \end{center}
  \end{minipage}
  \caption{ECR calculation procedure based on non-inferiority testing: (a) Multivariate non-inferiority
    testing based on IUT and (b) ECR calculation.}
  \label{fig:ecr_calculation}
\end{figure*}

Note that the original ECR framework \cite{shankaranarayanan:01} is
based on a single performance measure of web page delay and does not
provide any systematic comparison procedure taking into account the
statistical variability in measured
data. Fig. \ref{fig:ecr_calculation} shows the new procedure for
calculating ECR based on multivariate non-inferiority testing (i.e.,
non-inferiority testing extended by the IUT), where $N_{M}$ and
$N_{R}$ denote the number of performance measures adopted and the
number of values for $R$ (i.e., $R_{i}$'s) used for comparison,
respectively.

First, we need to obtain measures of the user-perceived performances
for representative applications/services (i.e.,
$M_{1},\cdots,M_{N_{M}}$) --- e.g., web page delay defined as the
average time taken to download an entire web page
\cite{shankaranarayanan:01} and the fraction of decodable frames per
group of pictures (GoP) (also called ``decodable frame rate (DFR)'')
for streaming video \cite{Ziviani:05} --- of the reference
architecture for the line rates of $R_{1},\cdots,R_{N_{R}}$ where
$R_{1}=min(R_{F},~R_{D})$, $R_{N_{R}}>0$, and $R_{i}>R_{j}$ for
$i<j$.\footnote{Note that the resolution of the ECR depends on the
  number of $R_{i}$'s (i.e., $N_{R}$) and proper choice of their
  values.}

Second, using the procedure shown in Fig. \ref{fig:ecr_calculation},
we find a value for the line rate of the reference architecture for
which the measures of the candidate architecture are statistically
non-inferior to those of the reference architecture. The null and the
alternative hypotheses of the non-inferiority testing for measure
$M_{i}$ (e.g., web page delay) are given by
\begin{equation}
  \left\{
    \begin{array}{l}
      H_{0}:~\mu_{i,C} - \mu_{i,R} \geq \delta_{i} \\
      H_{1}:~\mu_{i,C} - \mu_{i,R} < \delta_{i}
    \end{array}
  \right.
  \label{eq:hypotheses}
\end{equation}
where $\mu_{i,C}$ and $\mu_{i,R}$ denote population means of $M_{i}$
for the candidate and the reference architectures, respectively, and
$\delta_{i}$ represents the tolerance for the measure $M_{i}$. The
null hypothesis ($H_0$) is rejected if the limit of one-sided
confidence interval for the difference (i.e., $\mu_{i,C} - \mu_{i,R}$)
is less than the tolerance \cite{graphpad:equiv_test}. This means that
the candidate architecture is ``at least as good as'' the reference
architecture for the given measure $M_{i}$. Note that for each measure
$M_{i}$, we need to determine an appropriate tolerance value
($\delta_{i}$) and, if needed, change the hypotheses accordingly. For
example, we need to change the hypotheses for the DFR
of streaming video (i.e., the higher, the better) as follows:
\begin{equation}
  \left\{
    \begin{array}{l}
      H_{0}:~\mu_{i,C} - \mu_{i,R} \leq -\delta_{i} \\
      H_{1}:~\mu_{i,C} - \mu_{i,R} > -\delta_{i}
    \end{array}
  \right.
  \label{eq:alternative_hypotheses}
\end{equation}

\section{Virtual Test Bed for Experiments}
\label{sec:virtual_test_bed}
To support the new comparative analysis framework described in Section
\ref{sec:comparison_framework}, we need a flexible, yet
computationally powerful experimental platform; to carry out the
statistical hypothesis testing in the proposed ECR calculation
procedure, we need a sample of performance measures big enough to
compute a reliable test statistic. In case of simulation experiments,
this means that we have to repeat a simulation many times with a
different random number seed per run, which is quite challenging for
large-scale simulations.

The experimental platform also should be able to capture the
interaction of many traffic flows through a complete protocol stack,
which are generated either by actual users or, as a practical
alternative, based on user behavior models. A real, physical test bed
in this regard is hardly a viable option, considering a long cycle of
design and performance evaluation of new architectures and protocols,
at least at the stage of the clean-slate design.

In this regard, we decided to implement a virtual test bed composed of
detailed simulation models based on OMNeT++ \cite{Omnet++} and INET
framework \cite{INET} which provide models for end-user applications
as well as a complete TCP/IP protocol stack.\footnote{The implemented
  simulation models are available at
  ``http://github.com/kyeongsoo/inet-hnrl''.} Note that simulation
studies in optical networking area usually focus on the issues at the
physical and/or the data link layer only but neglect the issues at
higher layers due to the limit in computing power. The recent
introduction of high-performance computing (HPC) clusters and the
cloud computing \cite{Armbrust:09}, however, brings enormous computing
power at a much lower cost and, in case of the cloud computing,
on-demand basis; this enables researchers to carry out a series of
large-scale network simulations in a realistic environment, which was
neither practical nor economically feasible in the past. Specifically,
we are using Amazon elastic compute cloud (Amazon
EC2)\footnote{http://aws.amazon.com/ec2/} as a running platform for
the virtual test bed, while we are developing programs at a local HPC
cluster for ease of testing and debugging processes.

\begin{figure*}[!hbt]
  \centering
  \includegraphics*[angle=-90,width=.65\linewidth]{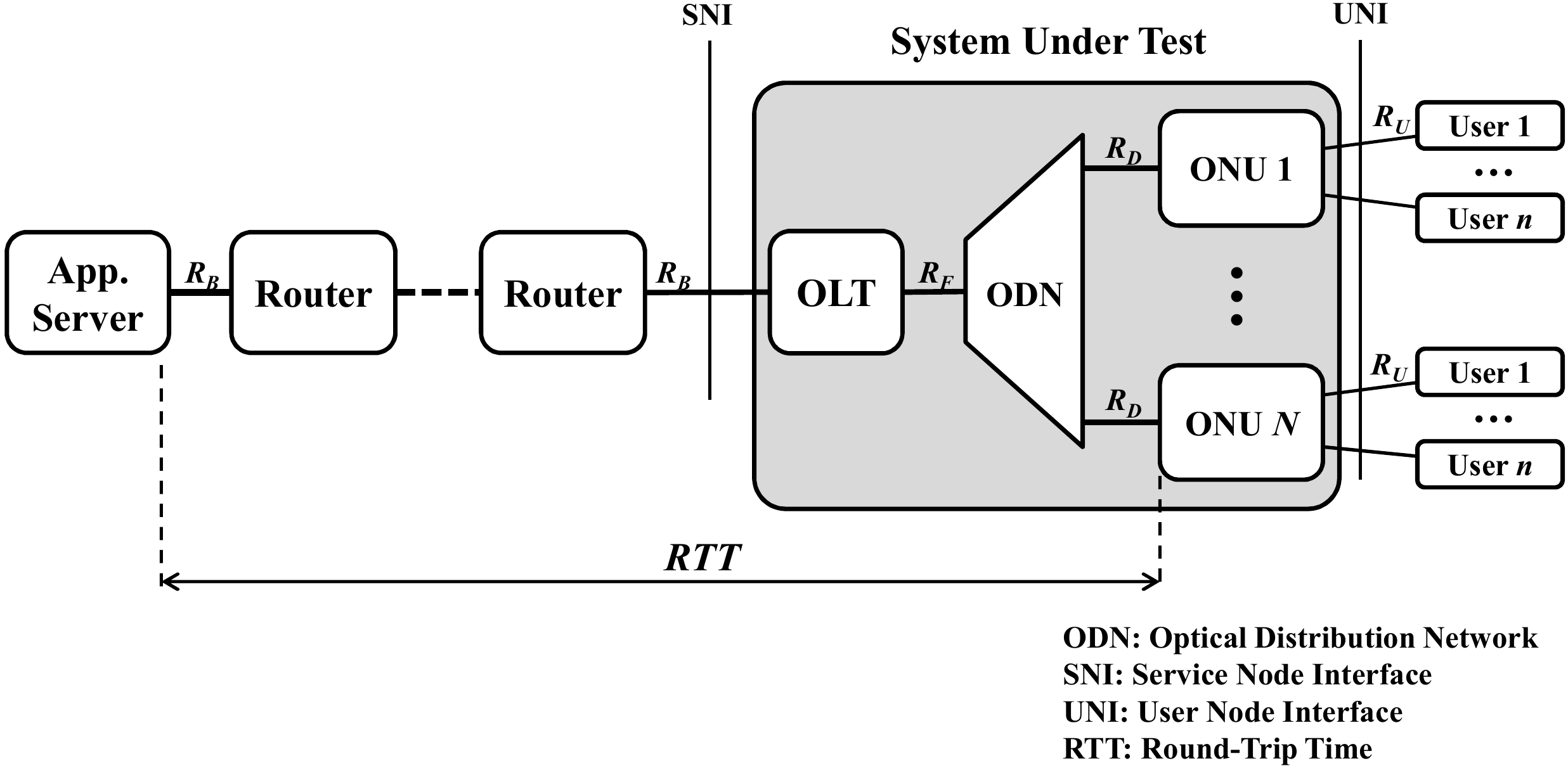}
  \caption{An overview of the virtual test bed.}
  \label{fig:virtual_testbed_overview}
\end{figure*}

Fig. \ref{fig:virtual_testbed_overview} shows the overview of the
virtual test bed where major building blocks are identified in
addition to the system under test between the service node interface
(SNI) and the UNI. For the reference architecture model, the OLT and
the ONU are implemented using a general IP router and an Ethernet
switch. For the candidate architectures, the hybrid TDM/WDM-PON under
the SUCCESS-HPON architecture \cite{Kim:05-1} has been already
implemented, and the implementation of 10G-EPON is currently underway.

\begin{figure}[!hbt]
  \centering
  \includegraphics*[angle=-90,width=\linewidth]{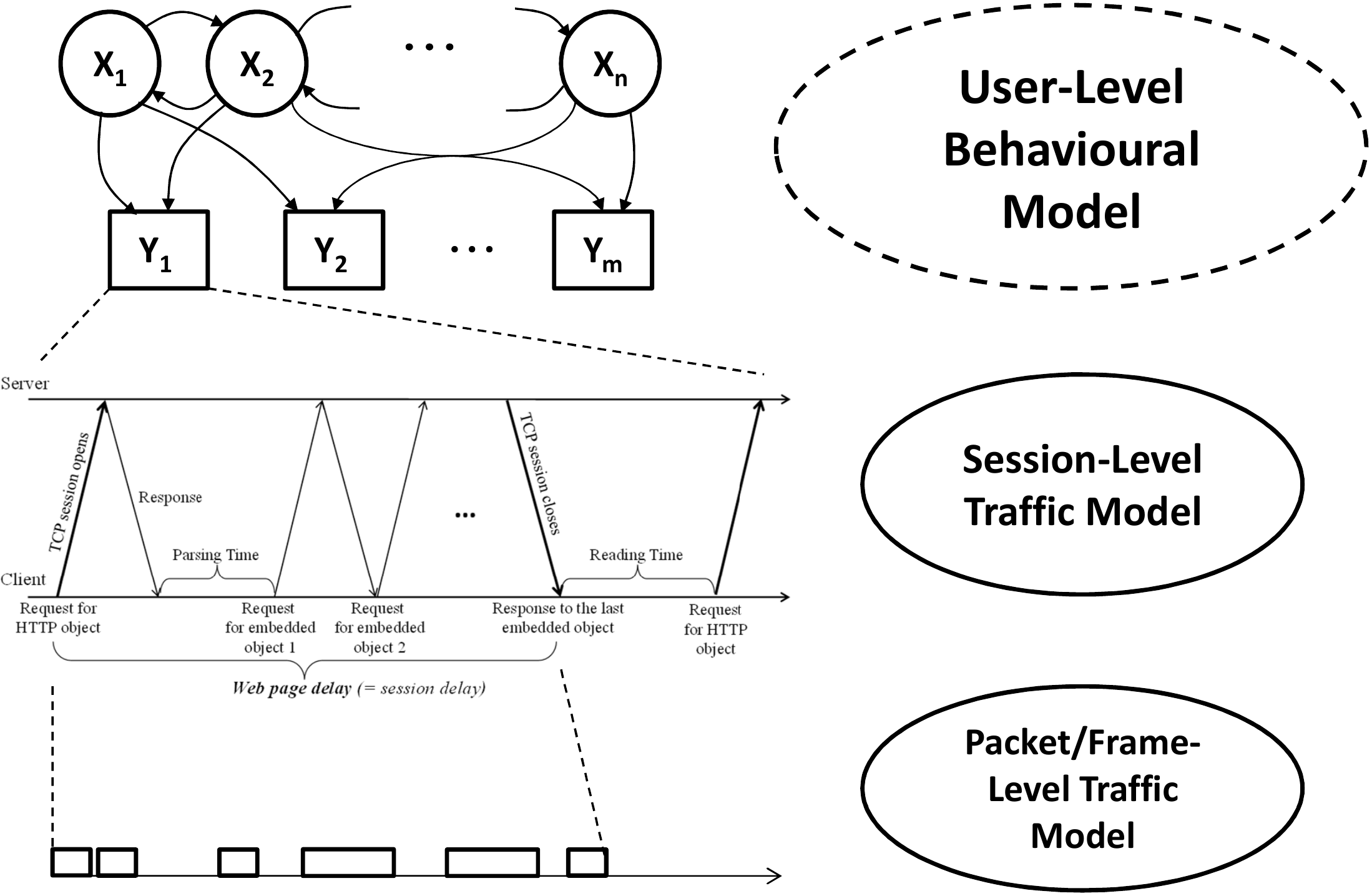}
  \caption{Multi-level traffic modelling and generation.}
  \label{fig:multi-level_traffic_modelling}
\end{figure}

Fig. \ref{fig:multi-level_traffic_modelling} shows our multi-level
approach to traffic modelling and generation in the virtual test bed;
as indicated by the dotted line in the figure, a user-level
behavioural model governing underlying applications/services is still
missing in the current implementation, while session-level and
packet-level models have already been implemented for HTTP, FTP, and
UDP steaming video \cite{Kim:11-1}.
We adapt the model proposed in \cite{Lee:07} for HTTP traffic
generation at the client side above TCP layer and without caching and
pipelining in a browser, while we use the FTP model from
\cite{cdma2000-evaluation:09} without any modification. For both the
traffic types, the virtual test bed provides a capability of measuring
session-level delay, throughput, and transfer rate as indirect measure
of user-perceived performances.

In addition to HTTP and FTP traffic, we also implemented a high-rate,
HD-TV-quality streaming video traffic for the virtual test bed, which
is considered one of killer applications for the NGOA. The implemented
traffic module can generate frames based on trace files from the ASU
video trace library \cite{Auwera:08-1}. As a measure of user-perceived
quality of video stream, we adopt the DFR which is defined as the
ratio of successfully decoded frames at a receiver to the total number
of frames sent by a video source \cite{Ziviani:05}: The larger the
value of DFR, the better the video quality perceived by the
end-user. For details of the implemented traffic models, readers are
referred to \cite{Kim:11-1,Kim:10-3}.

\begin{figure}[!hbt]
  \centering
  \includegraphics*[angle=-90,width=.9\linewidth]{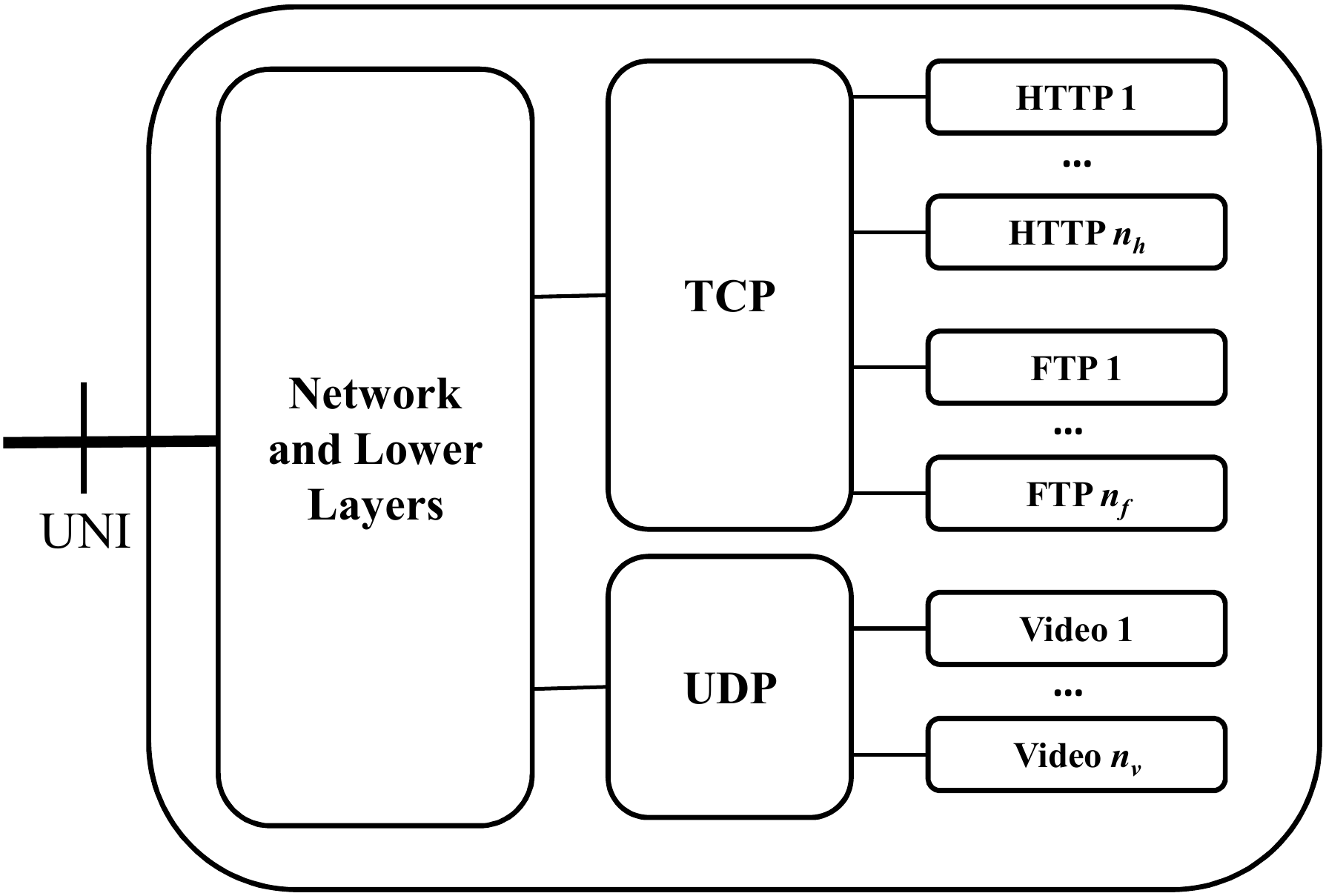}
  \caption{An end-user node (host) model.\label{fig:user_node_model}}
\end{figure}

Fig. \ref{fig:user_node_model} shows a model for an end-user node
which is connected to the ONU through UNI.
Currently, the number of traffic sessions (i.e., $n_h$, $n_f$, and
$n_v$) is static and configured at the beginning of a simulation
through an input file. Once we implement a user-level behavior model,
however, the number of traffic sessions will be dynamically controlled
by it during the simulation.
 
\subsection{Issues and Challenges}
\label{sec:issues_challenges}
Here are the lessons from our initial work reported in
\cite{Kim:10-3}: Because we need to run multiple simulations for given
parameters in order to compute a test statistic for the ECR
calculation, we repeated each simulation five times with different
random number seeds. Each simulation ran for 3 hours in simulation
time (much longer in real time), and the data were gathered after a
warmup period of 20 minutes; the warmup period should be long enough
to reduce the transient effects from the PON ranging procedure and
start-up delays introduced by streaming video encoding/decoding
processes as well as networking protocols like TCP.\footnote{We
  indirectly determined the warmup period of 20 minutes by
  investigating the total number of scheduled events in the
  future-event list; we observed that after 20 minutes, the number of
  scheduled events throughout the system goes into a steady state for
  all the simulations considered.} The total number of simulation runs
for the initial work is more than 1000 (i.e., 780 and 250 for the
reference and the shared architectures, respectively), and it took
several months to finish the whole simulations with a Linux HPC
cluster with 22 computing nodes, each with 8-GB memory and an 8-core
Intel Xeon CPU running at 2 GHz.

As we discussed before, the cloud computing could be a solution in
this regard; we can run 1000+ simulations simultaneously with the equal
number of virtual computers (or cores) in principle. To reduce the run
time of each individual simulation, however, we need another approach
on top of cloud computing: Parallelization. Fortunately, the OMNeT++
supports parallel simulation through message passing interface (MPI)
\cite{mpiforum} library, and we are currently extending the
implemented models for parallel simulation. Once the virtual test bed
is ready for parallel simulation, we can increase the number of
virtual computers (e.g., 2000 for 1000 simulations) to speed up run
times.

As for the user-level behavioral model currently missing, we will
build a demographic and behavioural user profile by focusing on groups
for initial exploration and surveying large-scale data
collection. Note that, because there is no NGOA network deployed now,
the use of demographic and behavioural profile obtained from the
survey for the architectural study is the only practical option. Then
we will build the user-level behavioural model governing underlying
application-level traffic models based on the developed profile, which
can capture the difference between business and residential users and
temporal aspects of end-user behaviours \cite{Kim:11-2}.

\section{Preliminary Results: Elasticity of Hybrid TDM/WDM-PON with
  Tunable Transceivers}
\label{sec:preliminary_results}
\begin{figure}[!hbt]
  \centering
  \includegraphics*[angle=-90,width=0.9\linewidth]{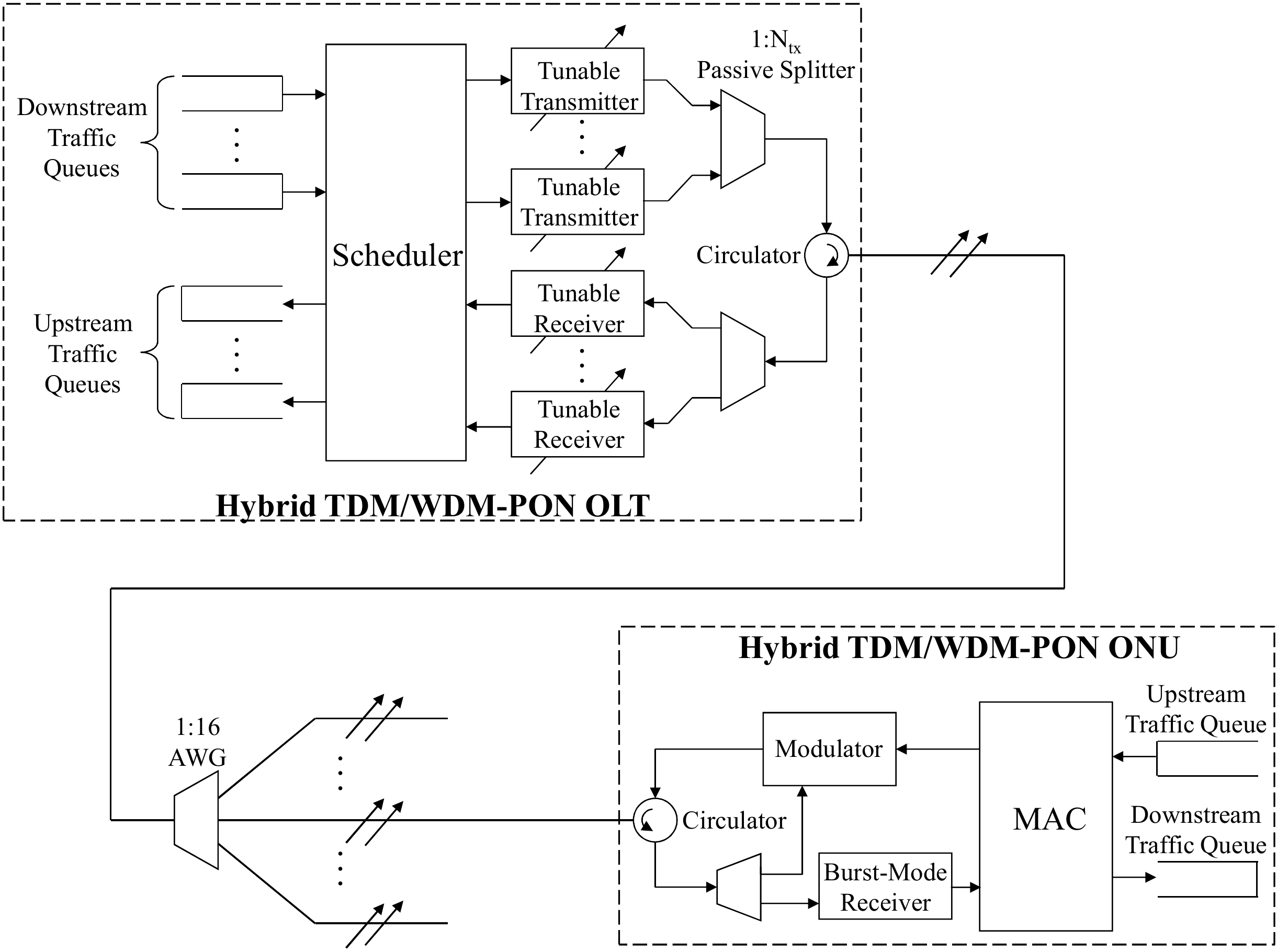}
  \caption{A block diagram of hybrid TDM/WDM-PON.}
  \label{fig:hybrid_pon}
\end{figure}

In this section, as a demonstration of the capability of the proposed
research framework, we present preliminary results of an ongoing study
of the elasticity of NGOA architectures. As of this writing, we have
carried out a simulation study of hybrid TDM/WDM-PON under
SUCCESS-HPON architecture with sequential scheduling with
schedule-time framing ($S^{3}F$) \cite{Kim:05-1}, whose block diagram
is shown in Fig. \ref{fig:hybrid_pon}.

The line rates $R_{B}$, $R_{D}$ and $R_{U}$ are set to 1 Tb/s, 10 Gb/s
and 10 Gb/s, respectively, $RTT$ to 10 ms, and $N$ (i.e., the number
of ONUs) to 16.

Because a user can interact with only one web page at a time, we set
the number of HTTP sessions to one, (i.e., $n_{h}=1$ in Fig.
\ref{fig:user_node_model}). The same is the case for a streaming video
(i.e, $n_{v}=1$). On the other hand, a user can run multiple FTP
sessions in the background. Therefore we set $n_{f}$ to 10, especially
to get a higher combined rate for 10-Gb/s access out of
well-established, lower-rate FTP parameters from 3GPP2
\cite{cdma2000-evaluation:09}.

\begin{figure}[!hbt]
  \begin{center}
    \includegraphics*[angle=-90,width=.9\linewidth]{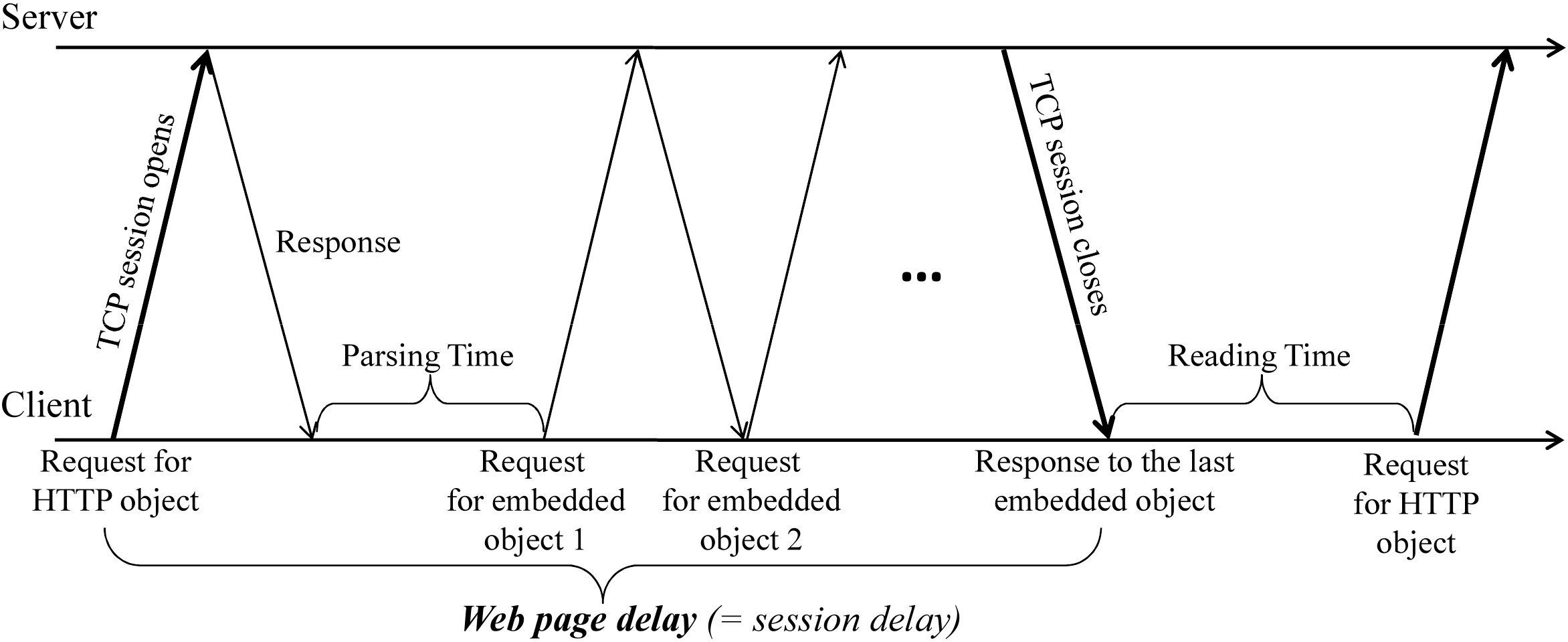}\\
    {\scriptsize (a)}\\
    \includegraphics*[angle=-90,width=.9\linewidth]{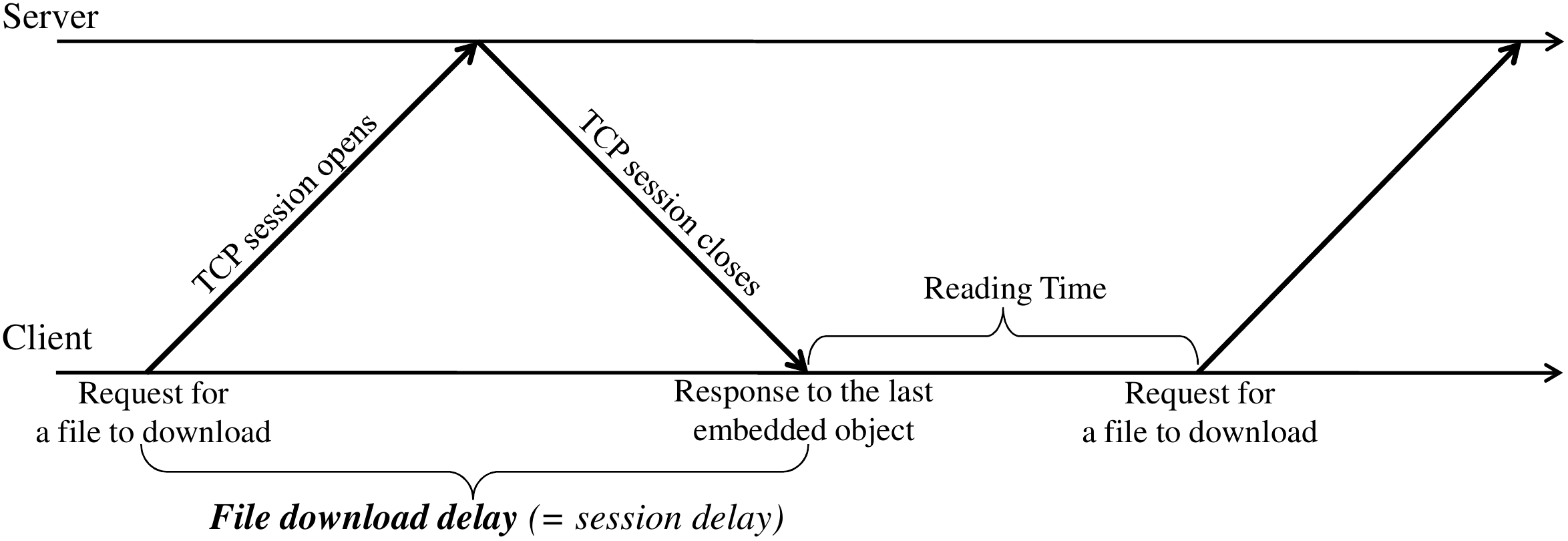}\\
    {\scriptsize (b)}
  \end{center}
  \caption{Traffic models for (a) HTTP and (b) FTP services.}
  \label{fig:http_ftp_traffic_models}
\end{figure}

Fig. \ref{fig:http_ftp_traffic_models} shows behavioral models for
HTTP and FTP traffic. The parameter values are summarized in Table
\ref{tbl:http_ftp_traffic_parameters}.

\begin{table}
  \centering
  \caption{Parameters for HTTP and FTP traffic models}
  \begin{tabular}{|l|l|} \hline
    Parameters/Measurements & Best Fit (Parameters) \\ \hline\hline
    \multicolumn{2}{|c|}{HTTP Model \cite{Lee:07}} \\ \hline
    HTML Object Size [Byte]: & Truncated Lognormal: \\
    Mean=11872, SD=38036, & $\mu$=7.90272, $\sigma$=1.7643, \\
    Max=2M & Max=2M \\ \hline
    Embedded Object Size [Byte]: & Truncated Lognormal: \\
    Mean =12460, SD=116050, & $\mu$=7.51384, $\sigma$=2.17454, \\
    Max=6M & Max=6M \\ \hline
    Number of Embedded Objects: & Gamma: \\
    Mean=5.07, Max=300 & $\kappa$=0.141385, $\theta$=40.3257 \\ \hline
    Parsing Time [sec]: & Truncated Lognormal: \\
    Mean=3.12, SD=14.21, & $\mu$=-1.24892, $\sigma$=2.08427, \\
    Max=300 & Max=300 \\ \hline
    Reading Time [sec]: & Lognormal: \\
    Mean=39.70, SD=324.92, & $\mu$=-0.495204, $\sigma$=2.7731 \\
    Max=10K & ~ \\ \hline
    Request Size [Byte]: & Uniform: \\
    Mean=318.59, SD=179.46 & $a$=0, $b$=700 \\ \hline\hline
    \multicolumn{2}{|c|}{FTP Model \cite{cdma2000-evaluation:09}} \\ \hline
    File Size [Byte]: & Truncated Lognormal: \\
    Mean=2M, SD=0.722M, & $\mu$=14.45, $\sigma$=0.35, \\
    Max=5M & Max=5M \\ \hline
    Reading Time [sec]: & Exponential: \\
    Mean=180 & $\lambda$=0.006 \\ \hline
    Request Size [Byte]: & Uniform: \\
    Mean=318.59, SD=179.46 & $a$=0, $b$=700 \\ \hline
  \end{tabular}
  \label{tbl:http_ftp_traffic_parameters}
\end{table}

\begin{table}
  \centering
  \caption{Overview of video traffic model}
  \begin{tabular}{|l|l|} \hline
    Property/Statistic & Value \\ \hline
    Video Clip & ``Terminator2 \cite{Auwera:08-1}'' \\
    Encoding & VBR-coded H.264/AVC \\
    Encoder & H.264 FRExt \\
    Duration & $\sim$10 min \\
    Frame Size & HD 1280x720p \\
    GoP Size & 12 \\
    Number of B Frames & 2 \\
    Quantizer & 10 \\
    Mean Frame Bit Rate & 28.6 Mb/s \\ \hline
  \end{tabular}
  \label{tbl:video_traffic_overview}
\end{table}

As for streaming video traffic, we use HD-TV-quality ``Terminator 2''
VBR-coded H.264/AVC clip from ASU video trace library
\cite{Auwera:08-1} as summarized in Table
\ref{tbl:video_traffic_overview}. Frames are encapsulated by real-time
transport protocol (RTP) and UDP before being carried in IP
packets. Considering that Ethernet frames are used in data link and
physical layers, the total overhead in this case is 66 octets
(=RTP(12)+UDP(8)+IP(20)+Ethernet(26)). The starting frame is selected
randomly from the trace at the beginning of simulation, and the whole
trace is cycled throughout the simulation.

\begin{figure}[!hbt]
  \begin{center}
    \includegraphics*[width=.8\linewidth,clip=true,trim=0mm 4mm 0mm 0mm]{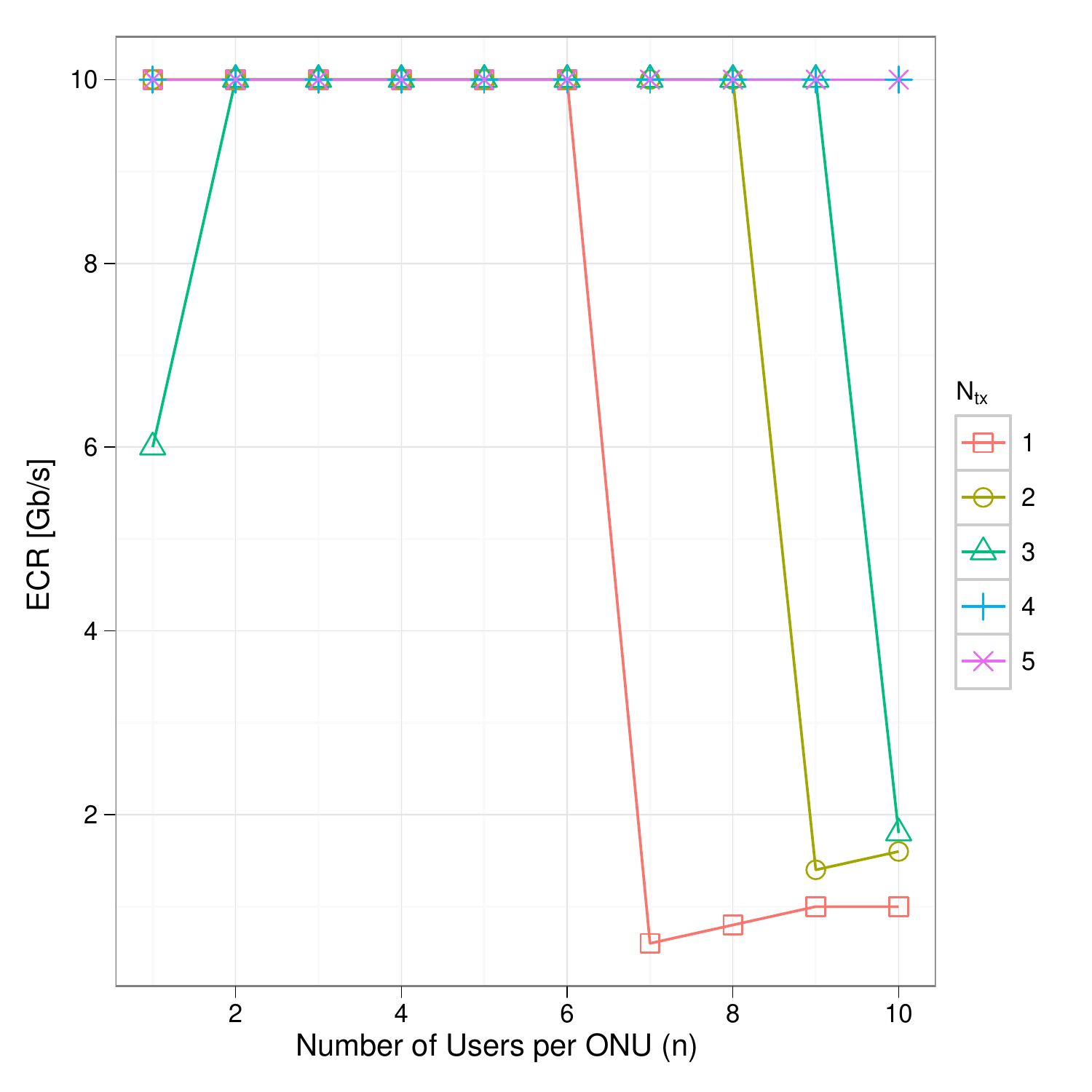}\\
    {\scriptsize (a)}\\
    \includegraphics*[width=.8\linewidth,clip=true,trim=0mm 4mm 0mm 0mm]{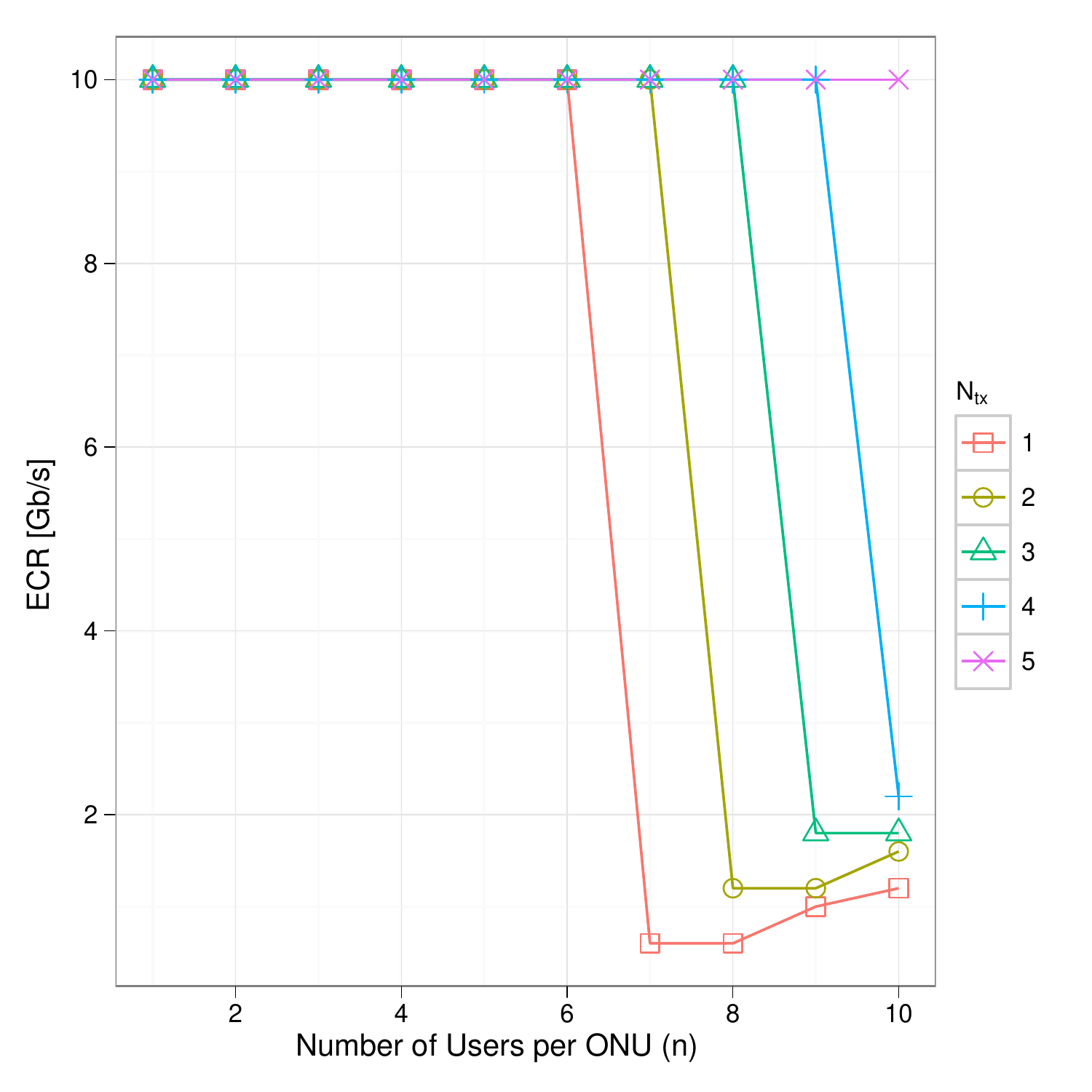}\\
    {\scriptsize (b)}
  \end{center}
  \caption{ECR of hybrid TDM/WDM-PON based on (a) web page delay and
(b) UDP streaming video decodable frame rate.}
  \label{fig:hybridpon_ecr}
\end{figure}

Based on the performance measures obtained from the simulation results
and the ECR calculation procedure described in Section
\ref{sec:comparison_framework}, we obtained ECRs of hybrid TDM/WDM-PON
as shown in Fig. \ref{fig:hybridpon_ecr}.\footnote{Due to space
  limitation, we provide various performance measures for all three
  types of traffic from the simulation in \cite{Kim:10-3}. Note that,
  because the results for FTP traffic was unreliable, especially when
  the load is high (probably due to the rather large number of
  sessions per user), we based the ECR calculation on the web page
  delay of HTTP traffic and the DFR of UDP streaming video only. In
  this case we can consider FTP traffic as background traffic, while
  the other two as test traffic.} We set the tolerance value to 10\%
of the sample mean of performance measure for the dedicated
architecture in non-inferiority testing with the significance level
($\alpha$) of 0.05. As expected, adding more transceivers greatly
improves the ECR as $n$ increases. For example, even with one more
transceiver (i.e., $N_{tx}=2$), we can achieve ECR of 10 Gb/s at
$n=7$, while the corresponding ECR with just one transceiver drops
below 1 Gb/s for both the results. On the other hand, it is remarkable
that the hybrid TDM/WDM-PON with just one transceiver can achieve ECR
of 10 Gb/s until $n$ reaches 6; when $n=6$, streaming video traffic
alone pushes about 180-Mb/s stream into ONU and 2.88-Gb/s multiplexed
stream into OLT (out of 16 ONUs). By the way, we found that the ECR
based on web page delay with $N_{tx}=3$ in
Fig. \ref{fig:hybridpon_ecr} (a) shows a rather strange value for
$n=1$. In fact, the sample mean of web page delay in this case is
2.6242 sec (with a confidence width of 0.19 sec) and just slightly
higher than those for $n=1$ which are in the range of [2.4958 sec,
2.5751 sec]. This anomaly is gone when we increase the tolerance value
to 20\% of the sample mean. Longer simulation run and a bigger sample
size could eliminate this anomaly. Also, the ECR values for
$N_{tx}=1,~2,~3$ suddenly increase once they reach the bottom. This is
because the performance measures are unreliable when the system is
highly overloaded.

\begin{figure}[!hbt]
  \begin{center}
    \includegraphics*[width=.75\linewidth,clip=true,trim=0mm 4mm 0mm 0mm]{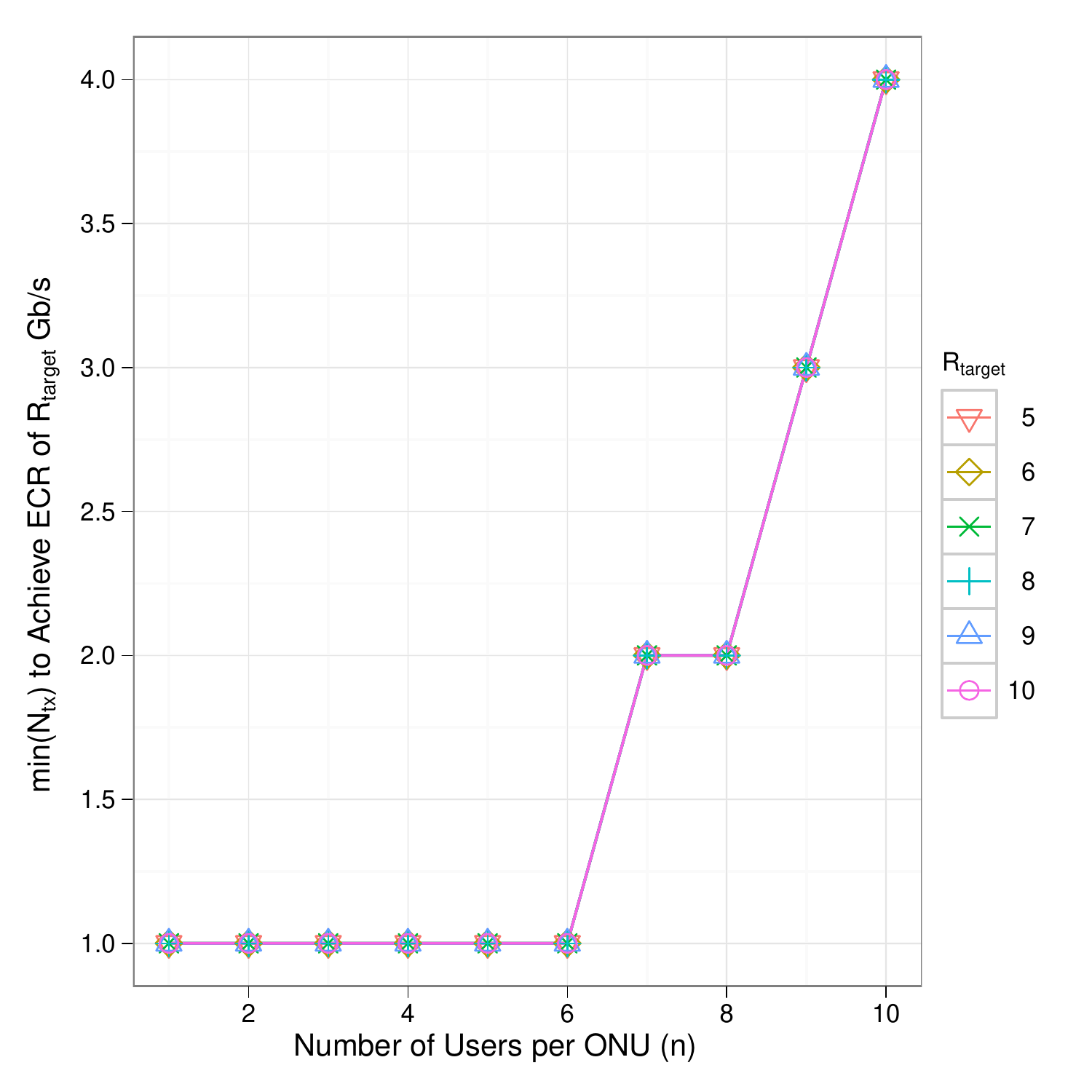}\\
    {\scriptsize (a)}\\
    \includegraphics*[width=.75\linewidth,clip=true,trim=0mm 4mm 0mm 0mm]{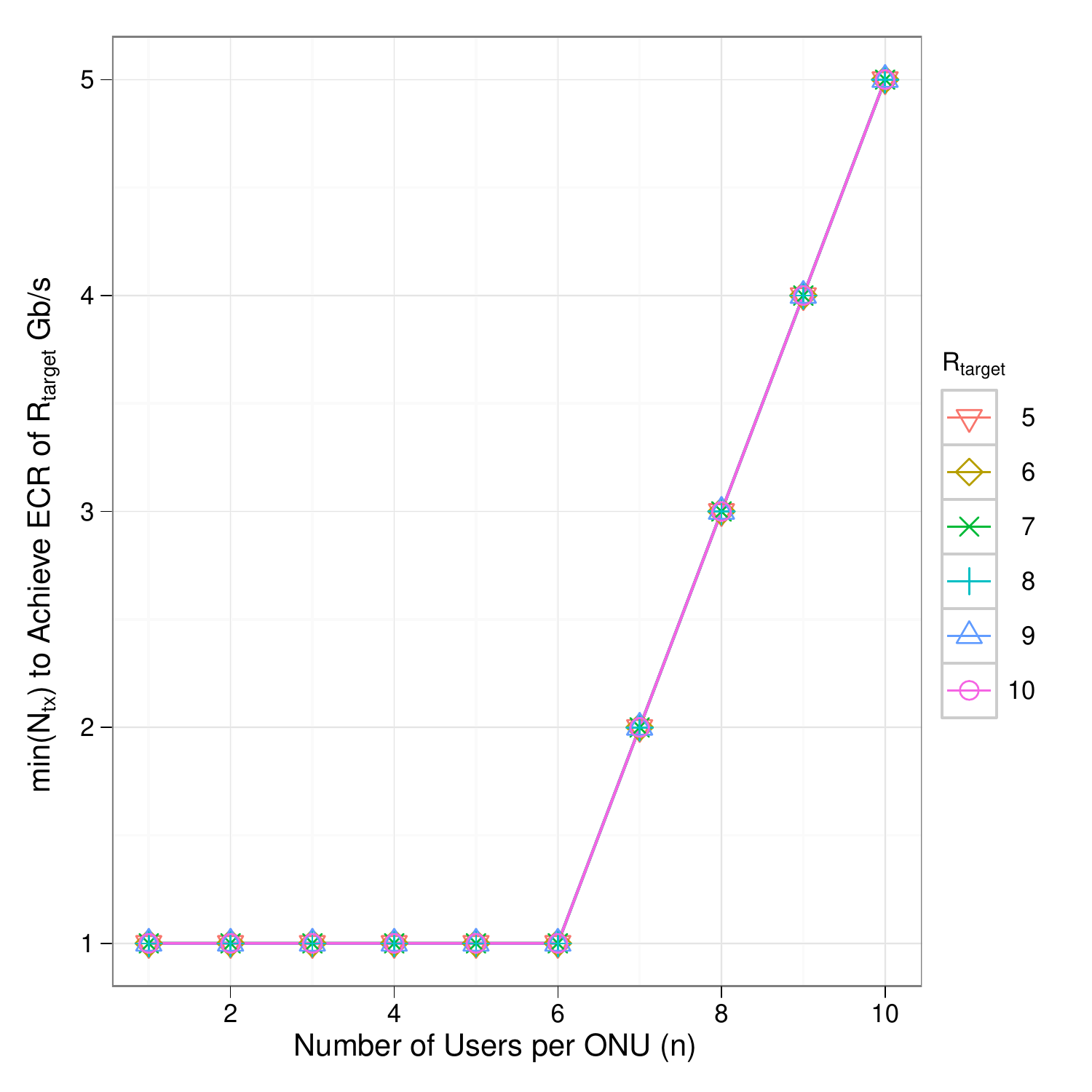}\\
    {\scriptsize (b)}
  \end{center}
  \caption{Minimum number of tunable transmitters ($min(N_{tx})$) to
    achieve ECR of $R_{target}$ in hybrid TDM/WDM-PON based on (a) web
    page delay and (b) UDP streaming video decodable frame rate.}
  \label{fig:hybridpon_min_ratio}
\end{figure}

The same results are plotted in a different way in
Fig. \ref{fig:hybridpon_min_ratio} which shows the minimum number of
tunable transmitters ($min(N_{tx})$) to achieve different ECR target
rates. $min(N_{tx})$ shows monotone increasing curve as $n$
increases. It is clear from the figure that we can achieve at least
ECR of 10 Gb/s with just one receiver until $n=6$. Note that, if we
consider both the performance measures together in ECR calculation
(i.e., through IUT), the resulting curves would follow those based on
the DFR of UDP streaming video in Fig. \ref{fig:hybridpon_min_ratio}
(b).

\section{Concluding Remarks}
\label{sec:conclusion}
We have reported the current status of our work on a new research
framework for the clean-slate design of NGOA architectures and
protocols, which is composed of the comparative analysis framework and
the virtual test bed for experiments. We described our proposed
comparative analysis framework based on the multivariate
non-inferiority testing and the ECR, which is capable of statistically
comparing multiple performance measures in an integrated way and
providing quantification of the effective bandwidth for users of a
candidate architecture with respect to a reference architecture. Then
we reviewed the current status of the virtual test bed implementation
and discussed the issues and challenges we met during initial studies
together with plans and strategies to address them in the next
version. To demonstrate the capability of the proposed research
framework, we also presented preliminary results from the ongoing
study of the elasticity of NGOA architectures, which show that the
hybrid TDM/WDM-PON can manage the same user-perceived performances as
those of the dedicated point-to-point architecture with the same line
rate by varying the number of tunable transceivers for a wide range of
traffic load.

Note that there is a major implication of the proposed research
framework: The way of comparing network architectures and protocols
and presenting their performances would be dramatically changed. For
instance, using the ECR as a reference (i.e., under the same ECR by
adjusting network configurations like the number of users), we can
objectively compare the issues of cost and energy efficiency of
candidate architectures, which is critical for the clean-slate design
of NGOA. In addition, the proposed research framework could greatly
help network service providers do proper dimensioning of their NGOA
before the actual deployment in the field, especially because the
notion of ECR is based on user-perceived performances.

Although the mix of traffic and their parameters in the proposed
research framework could be a good starting point, we need to further
refine them later based on the data from large-scale simulations
and/or, if possible, field trials in order to have a standard set of
traffic models for NGOA --- like those for wireless networking
\cite{wimax-sys-eval:08,cdma2000-evaluation:09}. Other important
topics not discussed in this paper are the inclusion of upstream
traffic (e.g., peer-to-peer applications) and the use of high-speed
variants of TCP \cite{Ha:06} in the virtual test bed.

\section{Acknowledgment}
The author would like to thank Dr Karin Ennser for fruitful
discussions in getting and presenting preliminary results in
Section \ref{sec:preliminary_results}. The author would also like to
thank Amazon for its support on this work through Amazon Web Services
(AWS) in Education Research Grant.


\end{document}